\documentclass[9pt,conference]{IEEEtran}
\IEEEoverridecommandlockouts

\usepackage{cite}
\usepackage{amsmath,amssymb,amsfonts}
\usepackage{algorithmic}
\usepackage{graphicx}
\usepackage{textcomp}
\usepackage{xcolor}
\usepackage{multirow}
\usepackage{svg}
\usepackage{soul}
\usepackage[normalem]{ulem}
\usepackage[numbers,sort&compress]{natbib}

\def\BibTeX{{\rm B\kern-.05em{\sc i\kern-.025em b}\kern-.08em
    T\kern-.1667em\lower.7ex\hbox{E}\kern-.125emX}}
\begin{document}

\title{Diffusion based Text-to-Music Generation\\ with Global and Local Text based Conditioning
\thanks{© 2025 IEEE.  Personal use of this material is permitted.  Permission from IEEE must be obtained for all other uses, in any current or future media, including reprinting/republishing this material for advertising or promotional purposes, creating new collective works, for resale or redistribution to servers or lists, or reuse of any copyrighted component of this work in other works.}
}

\author{\IEEEauthorblockN{Jisi Zhang}
\IEEEauthorblockA{\textit{Samsung R\&D}\\
\textit{Institute UK (SRUK)} \\ 
United Kingdom
}
\and
\IEEEauthorblockN{Pablo Peso Parada}
\IEEEauthorblockA{\textit{Samsung R\&D} \\
\textit{Institute UK (SRUK)} \\
United Kingdom
}
\and
\IEEEauthorblockN{Md Asif Jalal}
\IEEEauthorblockA{\textit{Samsung R\&D}\\\textit{Institute UK (SRUK)} \\ United Kingdom
}
\and
\IEEEauthorblockN{Karthikeyan Saravanan}
\IEEEauthorblockA{\textit{Samsung R\&D}\\\textit{Institute UK (SRUK)} \\ United Kingdom
}
}

\maketitle

\begin{abstract}
Diffusion based Text-To-Music (TTM) models generate music corresponding to text descriptions. Typically UNet based diffusion models condition on text embeddings generated from a pre-trained large language model or from a cross-modality audio-language representation model. This work proposes a diffusion based TTM, in which the UNet is conditioned on both (i) a uni-modal language model (e.g., T5) via cross-attention and (ii) a cross-modal audio-language representation model (e.g., CLAP) via  Feature-wise Linear  Modulation (FiLM). The diffusion model is trained to exploit both a local text representation from the T5 and a global representation from the CLAP. 
Furthermore, we propose modifications that extract both global and local representations from the T5 through pooling mechanisms that we call \textit{mean pooling} and \textit{self-attention pooling}. This approach mitigates the need for an additional encoder (e.g., CLAP) to extract a global representation, thereby reducing the number of model parameters. 
Our results show that incorporating the CLAP global embeddings to the T5 local embeddings enhances text adherence (KL=1.47) compared to a baseline model solely relying on the T5 local embeddings (KL=1.54). 
Alternatively, extracting global text embeddings directly from the T5 local embeddings through the proposed mean pooling approach yields superior generation quality (FAD=1.89) while exhibiting marginally inferior text adherence (KL=1.51) against the model conditioned on both CLAP and T5 text embeddings (FAD=1.94 and KL=1.47). Our proposed solution is not only efficient but also compact in terms of the number of parameters required.
\end{abstract}


\section{Introduction}
\label{sec:intro}
The field of generative AI has made significant progress recently leveraging large-scale computational resources, training data, and transformative AI advancements such as diffusion~\cite{Ho2020DenoisingDP} and transformers~\cite{Vaswani2017AttentionIA}. Diffusion based models have achieved the state-of-the-art generation performance in numerous areas such as text-to-image generation~\cite{Rombach2021HighResolutionIS}, natural language generation, video generation, text-to-audio generation~\cite{Liu2023AudioLDMTG,Liu2024audioldm2}, text-to-music generation~\cite{Evans2024FastTL}, speech synthesis~\cite{Popov2021GradTTSAD}, and so on.
Text-To-Music (TTM) models generate music conditioned over a given textual description of the desired music characteristics, such as type, style, genre, and instrumentation. Typically, a text description is processed through a pre-trained representation model~\cite{Radford2021LearningTV,Raffel2019ExploringTL,Huang2022MuLanAJ,laionclap2023} to create text embeddings, which are fed into a diffusion model that generates a latent space vector, further decoded (through models such as VAE and HiFi-GAN) into music~\cite{Liu2023AudioLDMTG,Liu2024audioldm2}. 

Contrastive Language-Image Pre-training (CLIP) based models have shown to outperform counterparts such as BERT for conditioning generative tasks~\cite{Devlin2019BERTPO}. 
The CLIP model is trained to generate a common representation between text and images~\cite{Radford2021LearningTV}. Other similar models such as the Contrastive Language-Audio Pretraining (CLAP)~\cite{laionclap2023} capture an audio-text representation~\cite{laionclap2023}. 
Conditioning through cross-modal models such as CLAP is particularly useful when high-quality annotated data is sparsely available during training~\cite{Agostinelli2023MusicLMGM,Liu2023AudioLDMTG}. For example, though training text-to-audio models would typically require 
text-audio annotated pairs, the audio embedding generated by CLAP could substitute the need for text inputs during training~\cite{Liu2023AudioLDMTG}. 

Improving the use of text encoders for conditioning is an active research area. Recently studies have shown that the CLAP text encoder suffers from an inability to capture temporal information~\cite{Wu2023AudioTextMD}, which are then resolved through conditioning over temporal-enhanced encoders~\cite{ copet2024simple,Liu2024audioldm2,Yuan2024TCLAPTC}.
Other works have also explored combining multiple text embeddings for conditioning diffusion models~\cite{Balaji2022eDiffITD, Liu2024audioldm2,Xue2024AuffusionLT}. Combining different types of embeddings in the text-to-image task has shown to alter generation outcomes~\cite{Balaji2022eDiffITD}.

In this paper, we categorize embeddings into \textit{global embeddings} when a single embedding vector represents the full input text and \textit{local embeddings} when multiple embedding vectors represent a given input text, e.g., one embedding per word.
A text-to-audio generation framework named Auffusion~\cite{Xue2024AuffusionLT} includes both global and local text embeddings during training and inference. The global and local embeddings are generated from a CLAP text encoder and a FLANT5 text encoder \cite{chung2024scaling} respectively. Both embeddings are then concatenated and fed into the diffusion UNet via a cross-attention mechanism. However, due to the different nature of the text embeddings created from different conditioning models, concatenation may not be the optimal approach for feature fusion. AudioLDM2~\cite{Liu2024audioldm2} employed GPT-2 to bridge various conditions generated from both a CLAP text encoder and a FLANT5 language model. Using an additional GPT-2 model introduces a large number of parameters to the model and increase the model complexity.

Although using a mixture of conditioning models can provide complementary information and improve the quality of generated music, there are no studies evaluating how different types of text encoder affect the generation quality.

\begin{figure*}[ht!]
  \centering
  \centerline{\includegraphics[width=14cm]{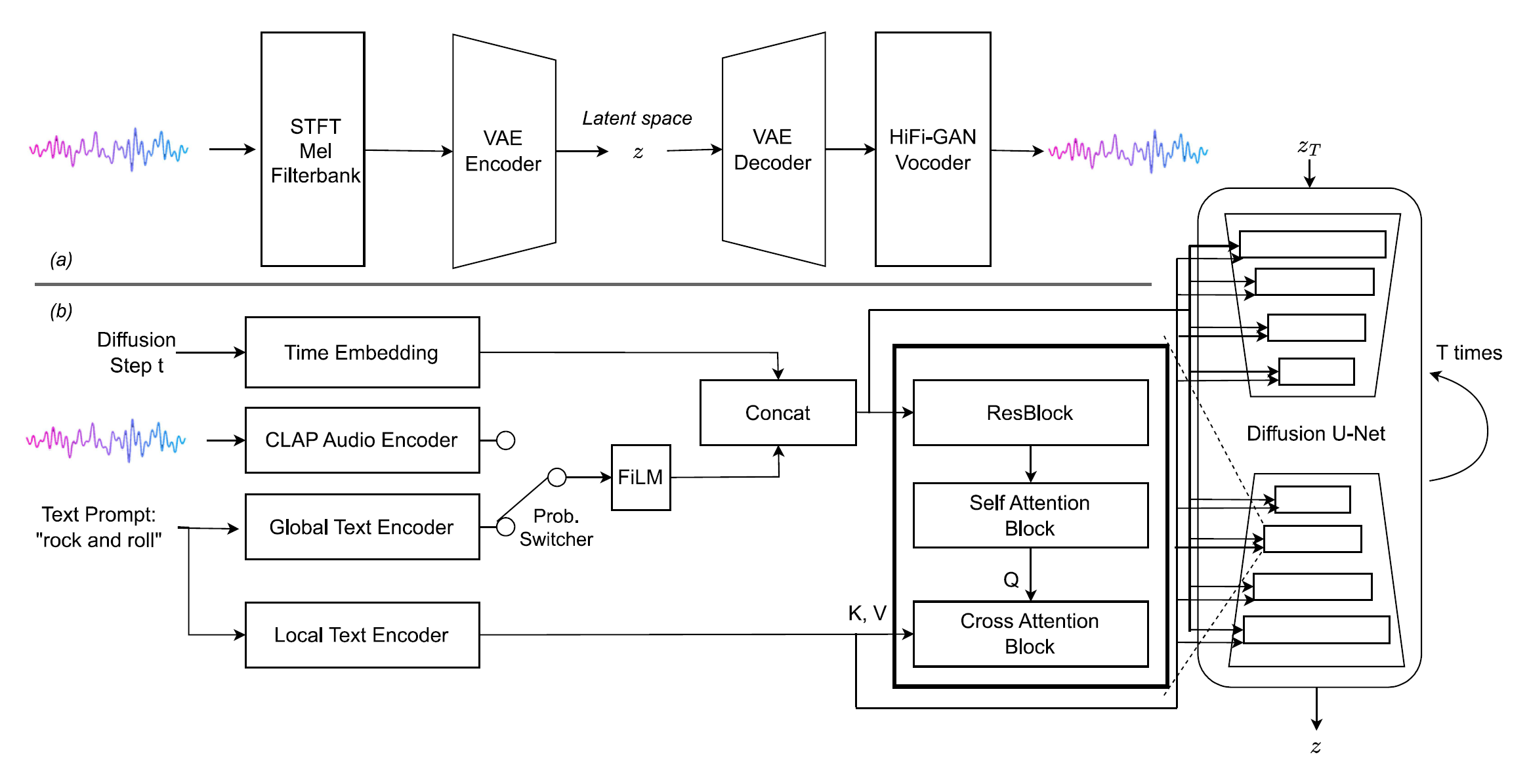}}
  
\caption{Overview of our text-to-music diffusion model architecture conditioned over multiple text encoders. (a) A continuous latent space $z$ is created from a VAE model. (b) A latent diffusion model is conditioned on text embeddings to generate the audio latent $z$ from a noisy version or white noise $z_T$ in T sampling steps. The CLAP audio encoder is only applicable when a CLAP model is used as a conditioner during training. K, V, and Q indicate the key, value, and query respectively in the attention mechanism.}
\label{fig:diffusion_framework}
\end{figure*}

The main contributions of this work are as follows:
\begin{itemize}
\item Analysis evaluating various text encoders that provide either local text embeddings or global text embeddings for conditioning the UNet in a diffusion based TTM model.
\item A novel conditioning mechanism that effectively takes both global and local text embeddings for the diffusion UNet. This mechanism injects global and local text embeddings at different levels in the diffusion UNet and does not need an additional model to fuse these two types of feature.
\item Two types of pooling methods, i.e. mean and self-attention pooling, are proposed in this work to extract global text embeddings directly from local text embeddings. It is shown that the mean pooling method can effectively extract a global text embedding to benefit the music generation performance in terms of quality (FAD=1.89) and text adherence (KL=1.51) without using an additional global text encoder. In comparison, the performance achieved with solely local text embeddings resulted in FAD=2.02 and KL=1.54.
\end{itemize}

\section{Method}

\subsection{Diffusion based generative model}
 This work builds upon latent diffusion models whose latent space is conditioned by pre-trained encoders~\cite{Rombach2021HighResolutionIS}. Following AudioLDM~\cite{Liu2023AudioLDMTG}, our setup includes a pre-trained autoencoder, a HiFi-GAN vocoder,  a text conditioner, and a UNet based latent diffusion model. The pre-trained autoencoder encodes mel-spectrograms into compressed latent space vectors. The HiFi-GAN vocoder constructs waveform signals from mel-spectorgrams~\cite{kong2020hifi}. The text conditioner converts text prompts to embeddings to condition the UNet based latent diffusion model to generate the music latent space. The UNet architecture is composed of encoder and decoder blocks, each of which is constructed using a ResNet layer and a spatial transformer layer \cite{Rombach2021HighResolutionIS}. The overall pipeline is shown in Fig \ref{fig:diffusion_framework}.

This diffusion model contains two processes, namely a diffusion process to gradually convert a clean latent representation $z_{0}$ to a noisy version $z_{T}$ and a reverse process to recover with multiple steps the data distribution corresponding to the clean latent representation from a noisy version or white noise.

\label{sec:method}

\subsection{Conditioning models}
Various text conditioning strategies are designed for the text-to-music generation.
Text-to-text transformer (T5) \cite{Raffel2019ExploringTL} is an encoder-decoder based language model trained on a generative span corruption pre-training task.
The T5 text encoder extracts local text embeddings. Most of the existing text conditioned music generation models rely on a cross-attention mechanism to inject local text embeddings in the generation process because of its flexibility and effectiveness \cite{copet2024simple,Liu2024audioldm2}. To enable efficient retrieval with sentence representation, the T5 model is later extended into a model that outputs a global text embedding, known as sentence-T5 \cite{ni2022sentence}.
The sentence-T5 explored three ways of extracting sentence embedding from the T5: (i) using the first token representation from the encoder; (ii) averaging all token representations of the encoder; (iii) using the first token representation from the decoder.

An alternative type of language model employs an encoder-only architecture and is usually trained on predicting masked tokens or predicting the next sentence in an unsupervised fashion \cite{Devlin2019BERTPO,Liu2019RoBERTaAR}. BERT \cite{Devlin2019BERTPO}, and RoBERTa \cite{Liu2019RoBERTaAR} are two examples of such pre-trained models. Global text embeddings can be extracted from a pre-trained model by taking the classification token representation, averaging embeddings from the pre-trained model \cite{Su2021WhiteningSR}, or using post-processing methods \cite{li2020sentence}. SimCSE \cite{Gao2021SimCSESC} applied a contrastive objective to pre-trained language models such as BERT or RoBERTa and greatly improves the global text embedding by regularising the pre-trained embeddings' anisotropic space to be more uniform.

CLAP is a language-audio multi-modality model that learns a joint global embedding space between text and audio through a contrastive learning framework \cite{laionclap2023}. A pre-trained text encoder and a pre-trained audio encoder separately encode text and audio input to text and audio embeddings, which are further mapped through text and audio projection layers to a joint space. Therefore the multi-modality model improves the audio representations from the natural language supervision and better aligns the text representation with the audio feature. Cross-modal representation models can be used to address the data scarcity issue of music-text pairs \cite{Agostinelli2023MusicLMGM}. Previous work uses embeddings computed from audio only as conditioning during training, and embeddings computed from text for inference \cite{Agostinelli2023MusicLMGM,Liu2023AudioLDMTG}

\subsection{Multiple text conditioning models}
To use multiple text encoders for conditioning, the diffusion UNet model is conditioned on a global text embedding and a separate local text embedding. The global text embedding is created from pre-trained models such as CLAP \cite{laionclap2023} or sentence-T5 \cite{ni2022sentence}. The local text embedding is created from a T5 model \cite{Raffel2019ExploringTL}.

In this work, given $y$ as the text prompt, the global text embedding $G_{y} \in \mathbb{R}^{d_{G}}$ is passed through a Feature-wise Linear Modulation (FiLM) \cite{Perez2017FiLMVR}, and concatenated with the time embedding, which is then used to bias the intermediate representation of the UNet. The local text encoder takes $y$ as input and outputs local text embedding $F_{y} \in \mathbb{R}^{M \times d_{F}}$. Due to the difference in nature between the global and local embedding, the local embedding is injected to the intermediate layers of the UNet via the cross-attention mechanism \cite{Vaswani2017AttentionIA}, $Attention(Q,K,V)=Softmax(\frac{QK^{T}}{\sqrt{d}}) V$, with
\begin{equation}
Q=\psi_{i}(z_{t}) W_{Q}^{(i)}, K=F_{y} W_{K}^{(i)}, V=F_{y} W_{V}^{(i)} 
\end{equation}
$\psi_{i}(z_{t})$ denotes the hidden representation of the $i$-th layer of the UNet and $z_{t}$ is a noisy latent representation at timestep $t$ corresponding to a pre-defined noise scheduler. $W_{Q}^{(i)}$, $W_{K}^{(i)}$, and $W_{V}^{(i)}$ are learnable projection matrices.

To avoid adding additional parameters from another text encoder, we propose a separate framework in which both the global and local text embedding are generated from the same text encoder. Specifically, when the T5 model is used as the text encoder, the global text embedding can be obtained from the local text embedding via either average pooling (\ref{equ:mean_pool}) or self-attention pooling (SAP) \cite{Safari2020SelfattentionEA}  (\ref{equ:sap_pool}).
\begin{equation}
\label{equ:mean_pool}
    G_{y}^{mean} = \frac{1}{M} \sum_{i=1}^{M} F_{y}^{(i)}
\end{equation}

\begin{equation}
\label{equ:sap_pool}
    G_{y}^{SAP} = Softmax(W_{SAP} F_{y}^{T})F_{y}
\end{equation}
where $W_{SAP} \in \mathbb{R}^{d_{F}}$ is a trainable parameter used to calculate the attention weights on the sequence of local text embedding.

The diffusion model $v_{\theta}$ is trained, using audio-text pairs, to optimise a v-objective function \cite{salimans2022progressive} defined as:
\begin{equation}
    v_{t} = \alpha_{t} \epsilon_{t} - \sigma_t z_{0}
\end{equation}
\begin{equation}
    L_{LDM} = \mathbb{E}_{\mathcal{E}(z),\epsilon \sim \mathcal{N}(0,1),t}\Bigl[\|v_{t}-v_{\theta}(z_t,t,G_{y},F_{y})\|_{2}^{2} \Bigr]
\end{equation}
where $\epsilon_{t} \in \mathcal{N}(0,1)$ and $\alpha_{t}, \sigma_t$ are noise scheduler parameters that typically define a signal-to-noise ratio at a timestep $t$ \cite{kingma2021variational}. The v-prediction is found to have a more stable training process than the epsilon parameterisation in our initial experiments thus we keep this v-prediction through this paper.

The pre-trained text encoders are kept frozen during the experiments. Classifier-free guidance (CFG) \cite{Ho2022ClassifierFreeDG} has been employed for guiding the diffusion model during training and generation processes. During training, both the global and local embeddings are randomly discarded with a probability of 0.1 to train the diffusion model for an unconditional generation task. During generation, the final noise estimation is a weighted combination of the conditional and unconditional output with a guidance scale $\omega$:
\begin{align*}
    \hat{v}_{\theta}(z_t,t,G_{y},F_{y})= &\omega v_{\theta}(z_t,t,\varnothing, \varnothing) \\
    &+ (1-\omega) v_{\theta}(z_t,t,G_{y},F_{y})
\end{align*}

\section{Experiment Setup}
\label{sec:exp_setup}

\subsection{Data}
Our dataset used for training music generation models contains publicly available datasets such as MTG \cite{Bogdanov2019TheMD}, Free Music
Archive (FMA) \cite{Defferrard2016FMAAD} and 10k high-quality commercial data from pond5~\cite{pond52023}. This work focused on generating instrument only music samples. Therefore, all the audio samples from both MTG and FMA are processed to filter out the vocal sound with a state-of-the-art music source separation, namely using the open source Hybrid Transformer for Music Source Separation (HT-Demucs) \cite{rouard2022hybrid}. MusicBench dataset \cite{Melechovsk2023MustangoTC} is used for validation. During training, the model is validated every 10k training steps to choose the checkpoint that yields the best FAD and KL scores. The proposed method is evaluated on MusicCaps benchmark \cite{Agostinelli2023MusicLMGM} to compare against prior work. MusicCaps contains 5.5k ten-second samples that are prepared by expert musicians. All the MusicCaps audio samples are processed with the HT-Demucs to keep only the instrumental part.

\subsection{Model configuration}
The VAE and HiFi-GAN vocoder used follow the same architectures as in AudioLDM~\cite{Liu2023AudioLDMTG} and are publicly available~\cite{AudioLDMtrain2024}.
The CLAP model uses a pre-trained checkpoint trained specifically for the music task released on the GitHub repository~\cite{CLAP2023}. The music generation models trained in this work generate 10 second audio tracks sampled at 16~kHz. The number of parameters for the diffusion UNet is approximately 300M. The diffusion models are trained with an AdamW \cite{Loshchilov2017DecoupledWD} optimiser with a learning rate of 1e-4 and 2000 steps of linear warming up without decay. The trained diffusion models are sampled using DDIM \cite{song2020denoising} for fast sampling. The CFG scale for the diffusion models trained in this work is set to 9.0. Other existing diffusion based music generation models \cite{Liu2024audioldm2,chen2024musicldm,Evans2024StableAO} in the literature are used for comparison and the parameters used for sampling are set to their default values. The generated output will be cropped to the first 10 seconds if it is longer than 10 seconds.

We use Fr\'echet Audio Distance (FAD) \cite{roblek2019fr} and Kullback-Leibler (KL) divergence metrics to measure the generated music quality with the AudioLDM evaluation toolkit~\cite{audioldmeval2023}. The FAD assesses the music quality based on similarity between the generated audio set and the target audio set in terms of the VGGish \cite{Hershey2016CNNAF} feature distribution. The KL divergence score measures the text adherence between the generated samples and the target samples, and is computed based on features extracted from a PANNs model \cite{Kong2019PANNsLP}.

\section{Results and Analysis}
\label{sec:result}
The overall performance of existing methods and our proposed methods is presented in Table \ref{tab:result_summary}, where the top part presents results using publicly available models and the bottom part presents the models trained on the training data mentioned in this work. We use the publicly available checkpoints from AudioLDM \cite{Liu2023AudioLDMTG}, AudioLDM2 \cite{Liu2024audioldm2}, MusicLDM \cite{chen2024musicldm}, and Stable Audio Open \cite{Evans2024StableAO} to generate music samples using MusicCaps captions and compute the FAD and KL scores against our processed test set. The small version models of AudioLDM and AudioLDM2 are selected for comparison since our developed diffusion model shares a similar model size. 

We re-train the diffusion network in the AudioLDM (CLAP) system with our own training data and achieve a better FAD score, but a worse KL score compared to the publicly available model. The improved FAD score suggests a better quality of the generated samples, which may be attributed to the high-quality commercial samples included in the training set. The degradation of KL could be attributed to the low quality text descriptions in the MTG and FMA data and the smaller quantity of total music data compared to the training data used for the publicly available AudioLDM model \cite{Liu2023AudioLDMTG}. Changing the text encoder from CLAP to T5 improves both FAD and KL, which matches the observation in other literature that a temporal-enhanced text encoder can improve the generation performance.

The proposed system (last row in Table \ref{tab:result_summary}) achieves better KL scores compared to models conditioned on either the CLAP global embeddings or the T5 local embeddings. This indicates that the diffusion network in the proposed system can effectively combine and exploit both global and local text embeddings extracted from the two different conditioner networks. The mixture of global and local text encoders achieves the best FAD score among existing models, surpassing the pre-trained AudioLDM2 that employs a GPT-2 model to bridge text encoder features of various modalities extracted from different condition models.
\begin{table}[t]
\centering
\caption{Objective evaluation of the proposed methods and existing models. The models on top use the publicly available pre-trained models. The models in the second half are trained on the training data used in this paper. Inference for the diffusion based models uses the DDIM sampler with 200 sampling steps.}
\label{tab:result_summary}
\begin{tabular}{c|cc|cc}
\hline
\multirow{2}{*}{\textbf{Model}} & \multicolumn{2}{c|}{\textbf{Text conditioner}} & \multirow{2}{*}{\textbf{FAD}} & \multirow{2}{*}{\textbf{KL}} \\
                                & \textbf{Global}        & \textbf{Local}        &                               &       \\ \hline
AudioLDM \cite{Liu2023AudioLDMTG}                 & CLAP                   & -                     & 2.60                       & 1.51  \\
AudioLDM2 \cite{Liu2024audioldm2}                 & CLAP                   & FLANT5-large               & 2.13                       & 1.36  \\
MusicLDM \cite{chen2024musicldm}& CLAP                   & -                     & 2.86                         & 1.41  \\
StableAudio \cite{Evans2024StableAO}   & -                 & T5-base               & 2.39                          & 1.56  \\ \hline
AudioLDM (retrain)        & CLAP                   & -                     & 2.35                       & 1.69     \\
AudioLDM (retrain)        & -                      & T5-base               & 2.02                          & 1.54  \\
LDM (ours)                & CLAP                   & T5-base               & 1.94                          & 1.47  \\
LDM (ours)                & CLAP                   & FLANT5-large          & 1.91                       & 1.47  \\\hline
\end{tabular}
\end{table}

\subsection{Ablation study}
In this section, we present our investigation into the performance of three different language models in providing the global text embeddings to the diffusion network: CLAP; Sentence-T5 \cite{ni2022sentence}; and SimCSE \cite{Gao2021SimCSESC}. The CLAP text encoder employs a pre-trained RoBERTa text encoder followed by a text projection layer, which produces the text embedding. For the Sentence-T5 model, a projection layer is applied to a pre-trained T5 base model to output a text representation with a fixed dimension and the model is fine-tuned with a contrastive loss. The SimCSE model fine-tunes a pre-trained RoBERTa model with both a dropout based unsupervised approach and a contrastive loss based supervised approach. Additionally the proposed mean pooling and self-attention pooling methods for extracting sentence embedding are included in the analysis. %

\begin{table}[t]
\centering
\caption{Performance of different strategies on providing global text embedding.}
\label{tab:global_emb}
\begin{tabular}{cc|c|cc}
\hline
\multicolumn{2}{c|}{\textbf{Text conditioner}} & \multirow{2}{*}{\textbf{\#Params}} & \multirow{2}{*}{\textbf{FAD}} & \multirow{2}{*}{\textbf{KL}} \\
\textbf{Global}        & \textbf{Local}        & &                               &       \\ \hline
CLAP                   & -                     & 497M & 2.35                          & 1.69        \\
-                      & T5-base               & 482M & 2.02                          & 1.54  \\
Sentence-T5            & T5-base               & 592M & 2.19                          & 1.54  \\
SimCSE                 & T5-base               & 606M & 2.94                          & 1.64  \\
CLAP                   & T5-base               & 606M & 1.94                          & 1.47  \\
Mean pooling            & T5-base               & 482M & 1.89                          & 1.51  \\
SAP                     & T5-base               & 482M & 2.05                          & 1.53  \\ \hline
\end{tabular}
\end{table}
The results are presented in Table \ref{tab:global_emb}. The first observation is that neither the fine-tuned Sentence-T5 nor the SimCSE as a global text embedding conditioner benefits music generation performance compared to only using the T5 model as a conditioner. We hypothesize that the Sentence-T5 and SimCSE models, having been fine-tuned on tasks unrelated to music generation, may not provide meaningful sentence representations from music generation textual prompts.

Averaging the local embeddings from the T5 model to create the global text embeddings improved both the FAD and KL performance compared to using the single T5 model. This indicates that the mean of the encoder outputs across all input tokens can naturally capture the sentence semantics and provide complementary information to the local embeddings. However, the sentence embedding extracted using the self-attention pooling shows no benefit to the performance. Our investigation showed that the trained attention pooling layer always attends to a narrow range of tokens that are dominant in the training dataset, which severely limits its expressiveness. This may require adding regularisation during training \cite{Gao2021SimCSESC} or post-processing strategies such as eliminating the dominant principal components or mapping emebeddings to an isotropic distribution~\cite{Su2021WhiteningSR}.

Although the best KL performance is achieved by the CLAP and T5 text encoder pair, it increases the model size and the inference time compared to using a single text encoder. Conversely, the mean pooling method for global text embedding extraction introduces no additional parameters and achieves better FAD and a slightly worse KL compared to the dual text encoder model.

\begin{table}[t]
\centering
\caption{Performance of different language models to provide local text embeddings (T5-base, FLANT5-large).}
\label{tab:local_emb}
\begin{tabular}{cc|c|cc}
\hline
\multicolumn{2}{c|}{\textbf{Text conditioner}} & \multirow{2}{*}{\textbf{\#Params}} & \multirow{2}{*}{\textbf{FAD}} & \multirow{2}{*}{\textbf{KL}} \\
\textbf{Global}        & \textbf{Local}        & &                               &       \\ \hline
CLAP                   & T5-base               & 606M & 1.94                          & 1.47  \\
Mean pooling            & T5-base               & 482M & 1.89                          & 1.51  \\
SAP                     & T5-base               & 482M & 2.05                          & 1.53  \\
CLAP                   & FLANT5-large           & 824M & 1.91                        & 1.47  \\
Mean pooling            & FLANT5-large          & 700M & 1.88                        & 1.50  \\ 
SAP                     & FLANT5-large          & 700M & 1.99                       & 1.50  \\ \hline
\end{tabular}
\end{table}

We further verify the benefits of the proposed pooling method via evaluation using two types of language model with different model capacity, namely T5-base and FLANT5-large. The FLANT5 model is an instruction-finetuned T5 model on a collection of tasks to improve model performance, especially to unseen tasks \cite{chung2024scaling}. The results are shown in Table \ref{tab:local_emb}. The first observation is that the dual text encoder model with FLANT5-large yields FAD improvement compared to the dual text encoder model which uses the T5-base model. We hypothesize that the improvement could come from the enhanced text representation from the FLANT5-large that has been fine-tuned on multiple tasks.
The proposed mean pooling method with the FLANT5-large achieved better FAD and KL compared to the mean pooling method which uses the T5-base model. 
This indicates that the pooling based method benefits from a language model that improves the global text representation, and can transfer over to improvements to music generation performance. This observation directs future research towards focus on improving the language model for the music generation task, including fine-tuning language models on the music generation task and incorporating audio understanding to powerful pre-trained language models \cite{Silva2023CoLLATOA}.

\section{Conclusions}
This work presents a thorough study on how both global and local text representations from various pre-trained language models affect the performance of diffusion based TTM models.
Firstly, we develop a simple and efficient TTM architecture that conditions global and local text representations at different levels of a diffusion UNet. The proposed model can effectively combine and exploit the two types of text representation to improve the generated music quality.
Next, we explore different pre-trained language models and pooling methods for obtaining the global text representations. Experiments show that, compared to a multiple text encoder conditioning model that obtains FAD=1.94 and KL=1.47, the mean pooling method achieves competitive music generation results (FAD=1.89 and KL=1.51). 
This highlights the parameter efficiency of the mean pooling technique and its effectiveness in extracting meaningful sentence semantics embeddings from a T5 encoder model, which can be leveraged to enhance the performance of music generation models.

\label{sec:conclusion}

\bibliographystyle{ieeetr}
\bibliography{refs}

\end{document}